\documentclass[runningheads]{llncs}
\usepackage[T1]{fontenc}
\usepackage{graphicx}
\usepackage{multirow}
\usepackage{booktabs}
\usepackage{makecell}
\usepackage[colorlinks,urlcolor=blue]{hyperref}
\usepackage{graphicx,subcaption}
\usepackage{fontawesome} 
\usepackage{comment}
\usepackage{float}
\usepackage[ruled,vlined]{algorithm2e}
\usepackage{fontawesome}
\usepackage{footnote}

\usepackage{xcolor,colortbl}
\definecolor{bg}{HTML}{f2f2f2}

\begin{document}
\title{MedMNIST-C: Comprehensive benchmark and improved classifier robustness by simulating realistic image corruptions}
%
%

\titlerunning{MedMNIST-C}
%
\author{Francesco Di Salvo \and Sebastian Doerrich \and Christian Ledig}
\authorrunning{F. Di Salvo, S. Doerrich, and C. Ledig}

%
\institute{xAILab Bamberg, University of Bamberg, Germany \\
\email{francesco.di-salvo@uni-bamberg.de}}
%


\maketitle              
%


\begin{abstract}

The integration of neural-network-based systems into clinical practice is limited by challenges related to domain generalization and robustness. The computer vision community established benchmarks such as ImageNet-C as a fundamental prerequisite to measure progress towards those challenges. Similar datasets are largely absent in the medical imaging community which lacks a comprehensive benchmark that spans across imaging modalities and applications. To address this gap, we create and open-source \texttt{MedMNIST-C}, a benchmark dataset based on the MedMNIST+ collection covering 12 datasets and 9 imaging modalities. We simulate task and modality-specific image corruptions of varying severity to comprehensively evaluate the robustness of established algorithms against real-world artifacts and distribution shifts. We further provide quantitative evidence that our simple-to-use artificial corruptions allow for highly performant, lightweight data augmentation to enhance model robustness. Unlike traditional, generic augmentation strategies, our approach leverages domain knowledge, exhibiting significantly higher robustness when compared to widely adopted methods. By introducing \texttt{MedMNIST-C} and open-sourcing the corresponding library allowing for targeted data augmentations, we contribute to the development of increasingly robust methods tailored to the challenges of medical imaging. The code is available at \href{https://github.com/francescodisalvo05/medmnistc-api}{github.com/francescodisalvo05/medmnistc-api}.


\keywords{Robustness \and Dataset \and Benchmark \and Data augmentation.}
\end{abstract}
%
%
%
\section{Introduction}
In the past decade, Deep Neural Networks (DNNs) have made significant improvements, achieving impressive results in various domains, including healthcare. As such the medical imaging community has witnessed a substantial rise of deep learning based decision-support systems, leading to significant advancements in fields such as radiology \cite{wong2019artificial}, dermatology \cite{maron2019systematic}, and pathology \cite{hekler2019pathologist}, often improving human-performances. However, traditional neural networks continue to face challenges, notably with adversarial samples and distribution shifts \cite{drenkow2021systematic}. These shifts commonly occur as they are caused by multiple factors including imaging machines (\textit{e.g.}, different vendors), post-processing techniques, patient characteristics, and acquisition protocols. 

\noindent While the natural imaging domain has advanced in addressing these issues through works such as the ImageNet-C benchmark \cite{hendrycks2019benchmarking}, the medical imaging domain has seen fragmented efforts, with individual works tackling specific modalities, as dermatology \cite{maron2021benchmark}, digital pathology \cite{zhang2022benchmarking,huang2023assessing}, blood microscopy \cite{zhang2020corruption}, and more. 
Additionally, they adopt diverse evaluation metrics, such as accuracy \cite{zhang2020corruption}, average corruption error (CE), corruption error of confidence (CEC) \cite{zhang2022benchmarking}, normalized CE, relative normalized CE \cite{huang2023assessing}, and their \textit{balanced} counterparts \cite{maron2021benchmark}. These works, while foundational, highlight the absence of a unified evaluation setting that can address the diversity of medical imaging modalities. 

\noindent Unanimously, they confirm the lack of robustness of widely adopted deep learning algorithms. To address model robustness in a simple way, we typically rely on generic augmentation techniques, such as MixUp \cite{zhang2017mixup} (blending examples and labels), CutMix \cite{yun2019cutmix} (interchanging image patches), RandAugment \cite{cubuk2020randaugment} (chain of random policies), and AugMix \cite{hendrycks2020augmix} (extends RandAugment through multiple chains and a consistency loss). However, those methods are not consistently effective across datasets. A recent work \cite{gao2023out} showed superior performance of \textit{targeted augmentation} over generic and domain-invariant ones. While showing promising results on distribution shifts for histopathological images, it lacks a wider validation for the medical domain. \newline
Motivated by the unresolved issues and the potential of targeted augmentation, our work carves out a new path. We first introduce \texttt{MedMNIST-C}, a comprehensive image corruption benchmark for $12$ datasets and $9$ imaging modalities, leveraging the publicly available MedMNIST+ \cite{medmnistv1} dataset collection (resolution: $224{\times}224$). The designed corruptions are dataset-specific and reflect the possible artifacts encountered during image acquisition and processing, simulating real-world artifacts or possible distribution shifts. Then, we employ the designed corruptions as custom \textit{targeted augmentations}. The proposed augmentations can be readily applied via our open-source APIs during training and meaningfully enhance the robustness of traditional domain-agnostic approaches. In summary: 

\begin{itemize}
	\item We \textit{design targeted image corruptions} to simulate real-world artifacts across the 12 datasets and 9 imaging modalities included in the MedMNIST+ dataset collection.
	\item We \textit{introduce a novel robustness benchmark}, bridging the widely utilized MedMNIST+ dataset with the well-established robustness evaluation framework of ImageNet-C. This provides a comprehensive framework to assess the robustness of algorithms across diverse imaging modalities.
	\item We \textit{evaluate the robustness of widely adopted deep learning architectures} using our benchmark, highlighting the need for enhancing model robustness. 
	\item We \textit{demonstrate the effectiveness of our targeted augmentations}, increasing the robustness through a simpler and more interpretable method, compared to traditional ones. 
	\item We \textit{publicly share our corrupted datasets and APIs} to facilitate the online, and dataset-specific augmentation of medical image analysis problems. 
\end{itemize}


\section{The MedMNIST-C dataset}


\subsection{Corruptions}

\label{section-corruptions}

\begin{figure*}[htp]
\centering
\includegraphics[width=1.0\linewidth]{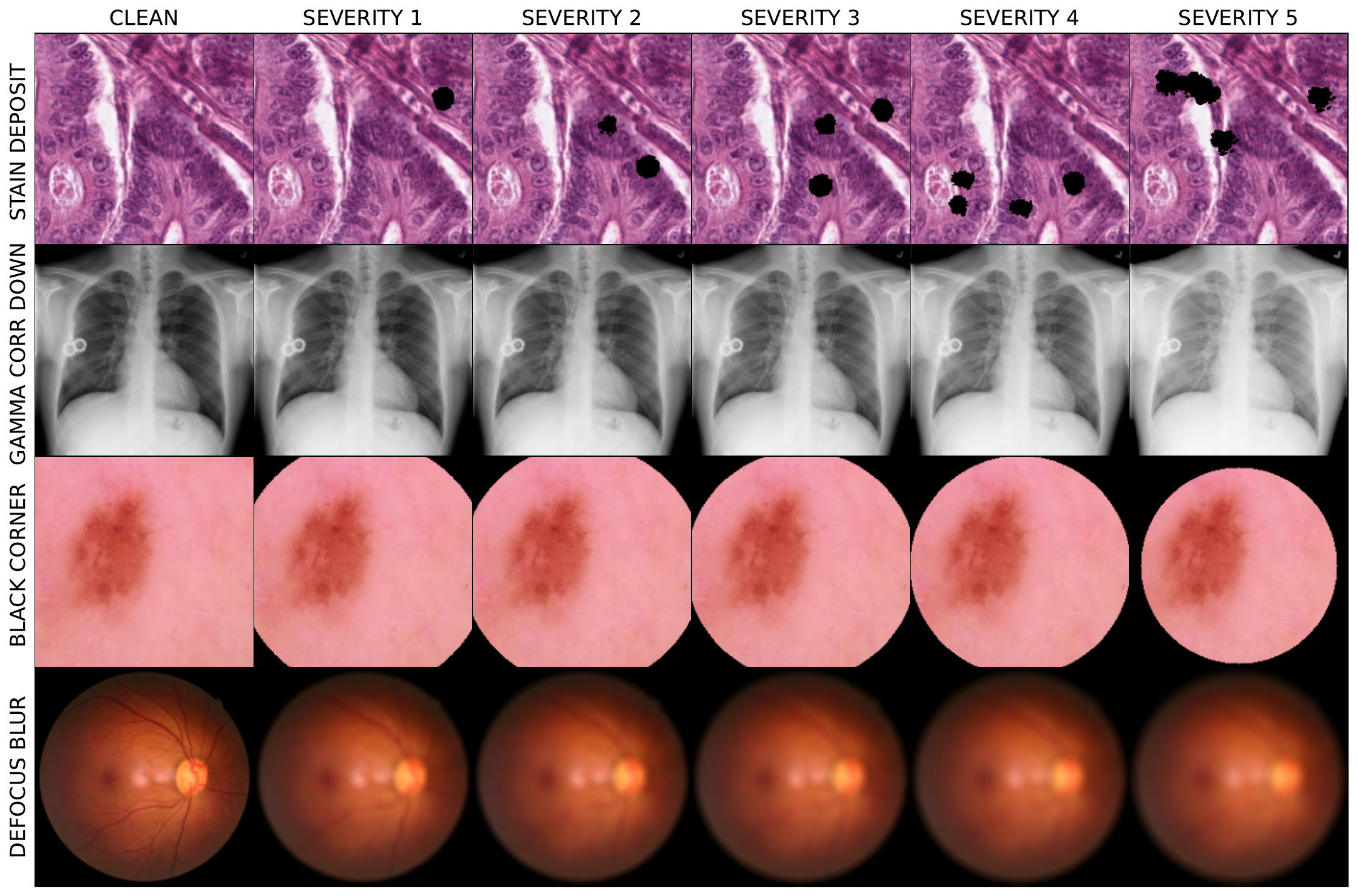}
\caption{\label{image-sample}Overview of four different corruptions applied (from top to bottom) to PathMNIST, ChestMNIST, DermaMNIST, and RetinaMNIST.}
\end{figure*}

This section summarizes the relevant literature and outlines the motivation for our chosen corruptions. As summarized in Table \ref{corr_overview}, we divided the corruptions into five categories, each with specific corruptions applicable to the respective MedMNIST+ (test) dataset. \texttt{Digital} corruptions are applied to any dataset and include \textit{JPEG} compression and \textit{pixelate}, the latter mimicking the effect of upsampling low-resolution images. The other categories are \texttt{noise}, \texttt{blur}, \texttt{color}, and \texttt{task-specific}. Following Hendrycks et al. \cite{hendrycks2019benchmarking} and subsequent medical imaging benchmarks \cite{maron2021benchmark,zhang2022benchmarking,huang2023assessing}, we evaluate model robustness across $5$ \textit{severity levels}. Therefore, we carefully reviewed, identified, and then described how those severity levels can be mapped to a broad set of medical imaging datasets, as illustrated in Figure \ref{image-sample}.


\begin{table}[h]
\centering
\def\arraystretch{1.25}\tabcolsep=2pt
\begin{tabular}{lccccc}
\toprule
                    & \multicolumn{5}{c}{\textbf{Corruption categories}}     \\  \cline{2-6}

\textbf{Dataset}    & \textbf{Digital}           
					& \textbf{Noise}             
					& \textbf{Blur}  
					& \textbf{Color}              
					& \textbf{Task-specific}      \\ \midrule

\makecell[l]{PathMNIST \\ BloodMNIST }           
					& \makecell{JPEG \\ Pixelate}
					& \makecell{-} 
					& \makecell{Defocus \\ Motion}
					& \makecell{Brightness\tiny{$[+|-]$} \\ Contrast\tiny{$[+|-]$} \\ Saturate}
					& \makecell{Stain deposit \\ Bubble} \\ \midrule

\makecell[l]{ChestMNIST \\
			 PneumoniaMNIST \\
			 OrganAMNIST  \\ 
			 OrganCMNIST \\
			 OrganSMNIST 
			 }           
		            & \makecell{JPEG \\ Pixelate}
					& \makecell{Gaussian \\ Speckle \\ Impulse \\ Shot} 
					& Gaussian 
					& \makecell{Brightness\tiny{$[+|-]$} \\ Contrast\tiny{$[+|-]$}} 
					& Gamma corr.\tiny{$[+|-]$}  \\ \midrule

DermaMNIST 			& \makecell{JPEG \\ Pixelate}
					& \makecell{Gaussian \\ Speckle \\Impulse \\Shot}
					& \makecell{Defocus \\ Motion \\ Zoom } 
					& \makecell{Brightness\tiny{$[+|-]$} \\ Contrast\tiny{$[+|-]$}}
					& \makecell{Black corner \\ Characters } \\ \midrule

RetinaMNIST         & \makecell{JPEG \\ Pixelate}                  
					& \makecell{Gaussian \\Speckle}
					& \makecell{Defocus \\ Motion}
					& \makecell{Brightness\tiny{$[-]$} \\Contrast\tiny{$[-]$}}
					& -  \\ \midrule

TissueMNIST         & \makecell{JPEG \\ Pixelate}                  
					& Impulse
					& Gaussian 
					& \makecell{Brightness\tiny{$[+|-]$} \\ Contrast\tiny{$[+|-]$}}
					& - \\ \midrule

OCTMNIST            & \makecell{JPEG \\ Pixelate}
					& Speckle 
					& \makecell{Defocus \\ Motion} 
					& \makecell{Contrast\tiny{$[-]$}}
					& - \\ \midrule
					
BreastMNIST         & \makecell{JPEG \\ Pixelate}
					& Speckle 
					& \makecell{Motion} 
					& \makecell{Brightness\tiny{$[+|-]$} \\ Contrast\tiny{$[-]$}}
					& - \\ \bottomrule
\end{tabular}
\caption{\label{corr_overview}Overview of the selected image corruptions, each applied at \textit{five} increasing severity levels. Note that the corruption hyperparameters are defined at dataset-level. The symbols {$[+|-]$} indicate whether the corruption intensity is increased, decreased,  or both, yielding separate corruptions.} 
\end{table}

\noindent \textbf{PathMNIST and BloodMNIST} \; \texttt{Pathology} and \texttt{blood cell microscopy} share similar imaging protocols, leading to common artifacts 
\cite{zhang2022benchmarking,zhang2020corruption,huang2023assessing}. These include \textit{stain deposits} and \textit{air bubbles}, along with \textit{defocus} and \textit{motion blur} during image acquisition. Variations in \textit{brightness}, \textit{contrast}, and \textit{saturation} are also prevalent, arising from different illumination and scanner conditions. \newline

\noindent \textbf{PneumoniaMNIST}\textbf{, ChestMNIST, and OrganMNIST (A,C,S)} \; \texttt{Chest X-Ray} and \texttt{abdominal CT} utilize X-ray radiation, with CT scans offering greater detail at higher radiation doses. Key corruptions \cite{navarro2021evaluating,momeny2021learning,islam2023robustness} include \textit{brightness} and \textit{contrast} variations, and \textit{Gaussian blur}. Additionally, X-ray images are particularly prone to \textit{Gaussian}, \textit{speckle}, \textit{impulse}, and \textit{shot} noise. Lastly, both modalities often undergo \textit{gamma correction} to adjust luminance. \newline 

\noindent \textbf{DermaMNIST} \; In \texttt{dermatoscopy}, skin lesions are examined using a dermatoscope. Supported by a comprehensive prior corruption benchmark \cite{maron2021benchmark} and further works like \cite{islam2023robustness}, we propose:  noise artifacts (\textit{Gaussian}, \textit{speckle}, \textit{impulse}, \textit{shot}), blurring effects (\textit{defocus}, \textit{motion}, \textit{zoom}), and color-based artifacts (\textit{brightness}, \textit{contrast}). Finally, we also include task-specific artifacts like \textit{black corners} from the dermatoscope and \textit{characters} from camera overlays. \newline

\noindent \textbf{RetinaMNIST} \; \texttt{Fundus photography} is essential for diagnosing retinal diseases. Here, image quality is affected by optimal \textit{brightness} and \textit{contrast}, influenced by lighting \cite{paulus2010automated}. To reflect diagnostic challenges, we also include \textit{Gaussian} and \textit{speckle} noise for electronic or sensor irregularities \cite{bala2021retinal}. Additionally, blurs like \textit{Gaussian} and \textit{defocus} are also frequent \cite{araujo2020dr}. \newline

\noindent \textbf{TissueMNIST} \; \texttt{High-throughput microscopy}, which is characterized by a high image acquisition rate, often experiences various image artifacts. Common issues include \textit{impulse noise} \cite{1628879} and \textit{blur} (Gaussian) \cite{mcelliott2023high}. Additionally, uneven illumination \cite{singh2014pipeline} is a frequent problem. This might affect the \textit{brightness} and \textit{contrast} of the images, resulting in inconsistent image quality.\newline

\noindent \textbf{OCTMNIST} \; \texttt{Optimal Coherence Tomography} (OCT) is an imaging modality providing high-resolution images of the retina. OCT images commonly exhibit several artifacts \cite{adabi2018overview,lian2018deblurring}, including \textit{speckle noise}, \textit{motion blur}, and \textit{defocus blur}. Notably, \textit{speckle noise} arises from the limited spatial-frequency bandwidth of the signal. This noise reduces the \textit{contrast} of the images, thereby degrading their quality and potentially affecting diagnostic accuracy. \newline

\noindent \textbf{BreastMNIST} \; \texttt{Breast ultrasound} uses sound waves to visualize breast tissue and detect abnormalities. Frequently observed artifacts \cite{sehgal2006review,huang2017breast} consist of \textit{speckle noise}, due to the scattering of ultrasound waves. Additionally, we observe variation in \textit{brightness}, low \textit{contrast}, and \textit{motion blur}. \newline


\subsection{Robustness measures}

\label{section_metric}

The MedMNIST+ datasets span four distinct tasks: binary classification, multi-class classification, multi-label (binary) classification, and ordinal regression. Following \cite{medmnistv1}, we treat ordinal regression tasks as multi-class classification problems, maintaining consistency across our evaluations. To develop our evaluation setting, we draw inspiration from the ImageNet-C benchmark \cite{hendrycks2019benchmarking}, which employs \textit{corruption error} and \textit{relative corruption error} as key metrics, subsequently adopted in \cite{huang2023assessing} in the context of digital pathology. Those metrics are rooted in the \textit{error} notation, intuitively defined as $1.0 - \textit{acc}$. Nevertheless, considering the diverse imbalance ratios across our datasets, we follow the approach of \cite{maron2021benchmark} for skin cancer, using the \textit{balanced error} (\textit{i.e.,} $1.0 - \textit{bacc}$). \newline

Thus, we first assess the clean balanced error, $\texttt{BE}_{clean}^{f}$, for each model $f$ using the respective MedMNIST+ test set. Next, we evaluate the balanced error, denoted as $\texttt{BE}_{s,c}^f$, for each corruption $c \in C_d$ and severity $s$ (from $1$ to $5$). Note that $C_d$ indicates the set of all corruptions associated with dataset $d$ (\textit{e.g.}, $C_{\tiny{blood}} = \{\textit{JPEG}, ..., \textit{brightness}+, \textit{brightness}-, ..., \textit{bubble}\}$). Then, we average across severities, normalizing those errors with AlexNet's errors, yielding $\texttt{BE}_c^f$. This step, formalized in Equation \ref{bce_cf}, accounts for the varying impacts of different corruptions on classification performance. Subsequently, we derive $\texttt{BE}^f$ by averaging across corruptions. 
 
\begin{equation}
	\texttt{BE}_c^f = \frac{\sum_{s=1}^5 \texttt{BE}_{s,c}^f}{\sum_{s=1}^5 \texttt{BE}_{s,c}^{AlexNet}}
	\label{bce_cf}
\end{equation}

\noindent Moreover, we measure the relative balanced error $\texttt{rBE}$ to assess the performance drop with respect to the clean test set. This is crucial to evaluate model's robustness, as we usually aim for minimal performance drop under distribution shifts. To do so, we simply extend Equation \ref{bce_cf} by subtracting the clean performance.\newline

\begin{equation}
	\texttt{rBE}_c^f = \frac{\sum_{s=1}^5 (\texttt{BE}_{s,c}^f - \texttt{BE}_{clean}^f)}{\sum_{s=1}^5 (\texttt{BE}_{s,c}^{AlexNet} - \texttt{BE}_{clean}^{AlexNet})}
	\label{rbce_cf}
\end{equation}

\noindent Finally, we average $\texttt{rBE}_c^f$ across all corruptions to derive $\texttt{rBE}^f$. For multi-label tasks, like ChestMNIST, metrics are averaged across labels as well.


\subsection{Corruption-API for Data Augmentation}

The data augmentation method employs the corruptions outlined in Section \ref{section-corruptions}. Specifically, for a given dataset $d$, we uniformly choose \textit{one} random corruption $c$ from the extended set $C_d'= C_d \, \cup \, \{\texttt{identity}\}$, where $C_d$ defines the corruptions associated with the imaging modality of dataset $d$, and \texttt{identity} represents no corruption. Upon selecting a specific corruption, we also randomly determine its severity. Note that each corruption's severity level is linked to distinct hyperparameters. For instance, with the \textit{brightness}$+$ corruption, the intensity ($i$) might range from $1.1$ at severity level $1$ to $1.9$ at level $5$. Thus, we randomly sample the corruption's hyperparameter using a uniform distribution that spans from the parameters defined for severity 1 and 5 (\textit{e.g.,} $i \sim U(1.1,1.9)$).



\section{Experimental results}

\textbf{Robustness} \; The first set of experiments aims to demonstrate the effect of our proposed target-specific corruptions on the performance of commonly used deep learning models. Specifically, we evaluate: AlexNet \cite{krizhevsky2012imagenet} (as normalizing baseline, following \cite{hendrycks2019benchmarking}), ResNet50 \cite{he2016deep}, DenseNet121 \cite{huang2017densely}, ViT-B/16 \cite{dosovitskiy2020image}, and VGG16 \cite{simonyan2014very}. All networks were initialized with ImageNet weights and trained on the respective MedMNIST+ training sets (\textit{cf.} Table \ref{table:augmentation_table} for number of training samples) for $100$ epochs with early stopping based on the respective validation set. The AdamW optimizer with a learning rate of $1e^{-4}$ was utilized, with a cosine annealing learning rate schedule. To evaluate the robustness we use the \texttt{BE} and the \texttt{rBE} described in Section \ref{section_metric}, averaging across three seed runs. \newline

\noindent Table \ref{robustness} presents a broad overview, averaging the model performance across all the proposed datasets. As expected \cite{paul2022vision}, while ViT-B/16 is the most robust one, VGG follows in terms of normalized performance (\texttt{BE}). However, despite achieving the best \textit{clean performance}, probably due to the highest number of parameters, it exhibits a severe performance drop (\texttt{rBE}). Notably, ResNet50 is the least robust one, followed by DenseNet121. This goes against the findings of the ImageNet-C benchmark. Thus, the training set size might play an important role. The table also shows that the degree of robustness is different among corruption types. For instance, \textit{task-specific} corruptions have the least impact on ResNet50, DenseNet121, and ViT. Also, while ViT exhibits higher robustness to all corruptions, the gap is even more pronounced on \textit{noise} and \textit{digital} artifacts. 

\begin{table}[h]
\centering
\begin{tabular}{lcccccccccc}
\toprule

& & M+ & \multicolumn{3}{c}{\texttt{M-C}} & \multicolumn{5}{c}{\textbf{BE} $\downarrow$} \\ 

\cmidrule(r){3-3}
\cmidrule(r){4-6}
\cmidrule(r){7-11}

\textbf{Arch} & \textbf{\#Par} 
			  & \cellcolor{bg}\textbf{bACC $\uparrow$}
					  & \cellcolor{bg}\textbf{bACC $\uparrow$} 
                      & \cellcolor{bg}\textbf{rBE $\downarrow$} 
                      & \cellcolor{bg}\textbf{BE $\downarrow$} 
					  & \textbf{Digital}
					  & \textbf{Noise}
					  & \textbf{Blur}
					  & \textbf{Color}
					  & \textbf{TS} 
                      \\ \midrule
AlexNet & 62.3 M 
			 & \cellcolor{bg} 78.7 
			 & \cellcolor{bg} 62.9 
			 & \cellcolor{bg} 100.0 
			 & \cellcolor{bg} 100.0 
			 & 100 
			 & 100 
			 & 100 
			 & 100 
			 & 100 
\\ 
R.Net50 & 25.6 M 
			 & \cellcolor{bg} 75.4 
			 & \cellcolor{bg} 56.2 
			 & \cellcolor{bg} 166.1 
			 & \cellcolor{bg} 131.5 
			 & 177 
			 & 110 
			 & 123 
			 & 148 
			 & 95 
\\ 
D.Net121 & 8 M 
			 & \cellcolor{bg} 79.8 
			 & \cellcolor{bg} 59.4 
			 & \cellcolor{bg} 148.4 
			 & \cellcolor{bg} 114.8 
			 & 145 
			 & 124 
			 & 100 
			 & 124 
			 & 78 
\\ 
VGG16 & 138.4 M 
			 & \cellcolor{bg} 80.5 
			 & \cellcolor{bg} 65.9 
			 & \cellcolor{bg} 114.0 
			 & \cellcolor{bg} 93.0 
			 & 128 
			 & 87 
			 & 91 
			 & 84 
			 & 80 
\\ 
ViT-B & 86.6 
			 & \cellcolor{bg} 78.9 
			 & \cellcolor{bg} 72.0 
			 & \cellcolor{bg} 59.9 
			 & \cellcolor{bg} 76.3 
			 & 74 
			 & 50 
			 & 77 
			 & 80 
			 & 71 
\\ 
\bottomrule
\end{tabular}

\caption{\label{robustness} Balanced accuracy, \texttt{BE}, and \texttt{rBE}, averaged across all $12$ datasets included in our benchmark, spanning from low-data settings (${<}1{,}000$ train samples, \textit{e.g.,} BreastMNIST) to large-data ones (${>}100{,}000$ train samples, \textit{e.g.,} TissueMNIST). Note that {\#Par} refers to the number of parameters, M${+}$ refers to the clean MedMNIST${+}$ test set, and \texttt{M-C} refers to our corrupted test set.}
\end{table}

\vspace{-1mm}

\noindent \textbf{Data augmentation} \; The second experiment aims to evaluate our augmentation method against reference ones, on the corrupted test sets, utilizing a ResNet18 for consistency. We benchmark our method against generic augmentation approaches: MixUp ($\alpha=0.2$), CutMix ($\alpha=1.0$), and RandAugment ($k=1$), with $\alpha$ values chosen from recommended ranges \cite{gao2023out}. RandAugment, sharing similarities with our approach through augmentations like \textit{color, contrast, brightness,} and \textit{sharpness}, differs by employing domain-agnostic intensity ranges. To fairly compare our method, we limit RandAugment to a single corruption at a time ($k=1$). For the same reason, we exclude from our comparison AugMix, which builds upon RandAugment with multiple chains. In future work, the extension of our targeted augmentations along multiple chains could be explored. While the generic augmentations are employed once and tested on the corrupted test sets, we apply a double evaluation for our method. Initially, considered as a performance baseline, we train and test using a $k$-fold cross-validation, with folds organized by corruption types: \texttt{digital}, \texttt{noise}, \texttt{blur}, \texttt{color}, and \texttt{task-specific}. Thus, by training and evaluation with semantically different corruptions, we carefully ensure no overfitting. Secondly, we evaluate using all corruptions, similar to competing methods. Focusing on absolute gains, we report the Area Under the Curve (AUC), averaging across three seed runs. \newline 

\noindent Table \ref{table:augmentation_table} shows that, even under a $k$-fold cross-validation method, our augmentation approach achieves the top average AUC gain across all 12 datasets. Notably, our augmentation excels particularly in smaller datasets, those with fewer than $10{,}000$ samples, which also tend to have lower initial AUC values. In contrast, without the $k$-fold settings, we clearly observe significant improvements. While this might seem as an overestimate of performance gains, it pinpoints a sweet spot for our method's effectiveness, marking substantial enhancements over generic domain-agnostic approaches. This underscores the importance of integrating domain knowledge into data augmentation strategies.


\begin{table}[h]
\centering
\begin{tabular}{lcccccccccccc|c}
\toprule

\textbf{Dataset}	  & \textbf{Br}
					  & \textbf{Re}
					  & \textbf{Pn}
					  & \textbf{De}
					  & \textbf{Bl}
					  & \textbf{OrC}
					  & \textbf{OrS}
					  & \textbf{OrA}
					  &	\textbf{Ch} 
					  & \textbf{Pa}
					  & \textbf{OCT} 
					  & \textbf{Ti} 
					  & \textbf{AVG} \\

\textbf{\#Train}  	  & \tiny{546}
					  & \tiny{1{,}080}
					  &	\tiny{5{,}856}
					  & \tiny{7{,}007}
					  & \tiny{11{,}959}
					  & \tiny{12{,}975}
					  & \tiny{13{,}932}
					  & \tiny{34{,}561}
					  & \tiny{78{,}468}
					  & \tiny{89{,}996}
					  & \tiny{97{,}477}
					  & \tiny{165{,}466} \\  
					  
\textbf{\#Classes} 	  & \tiny{2}
					  & \tiny{5}
					  &	\tiny{2}
					  & \tiny{7}
					  & \tiny{8}
					  & \tiny{11}
					  & \tiny{11}
					  & \tiny{11}
					  & \tiny{14}
					  & \tiny{9}
					  & \tiny{4}
					  & \tiny{8} \\  \midrule
No Aug.
			& 73.4
			& 70.8
			& 84.7
			& 78.1
			& 97.8
			& 91.5
			& 89.9
			& 92.0
			& 66.2
			& 92.5
			& 89.3
			& 82.8
			 & 84.1
 \\ 

\midrule 

CutMix
			& 1.2
			& 2.4
			& 7.1
			& -0.7
			& -0.8
			& 1.1
			& 2.2
			& 0.4
			& (0) 
			& -1.5
			& -0.8
			& 0.4
			 & 1.0
 \\ 

MixUp
			& 1.6
			& 0.8
			& 5.1
			& 3.0
			& 0.2
			 & \textbf{5.0}
			& 2.9
			 & \textbf{4.8}
			& (0) 
			& 3.4
			& 2.0
			 & \textbf{1.0}
			 & 2.7
 \\ 

RandAug.
			& 6.7
			& \textbf{3.2}
			 & \textbf{8.7}
			& 4.1
			 & \textbf{1.3}
			& 4.0
			 & \textbf{4.8}
			& 2.2
			& 3.2
			 & \textbf{4.3}
			& 3.4
			& 0.8
			 & 3.9
 \\ 

\midrule 

Ours ($k$-F.)
			 & \textbf{8.5}
			 & \textbf{3.2}
			& 7.9
			 & \textbf{6.8}
			& 0.4
			& 2.2
			& 3.9
			& 2.1
			 & \textbf{4.7}
			& 3.2
			 & \textbf{4.5}
			& 0.9
			 & 4.0
 \\ 

Ours
			 & \textbf{12.7}
			 & \textbf{8.5}
			 & \textbf{12.6}
			 & \textbf{16.3}
			 & \textbf{2.1}
			 & \textbf{7.6}
			 & \textbf{7.8}
			 & \textbf{7.4}
			 & \textbf{10.1}
			 & \textbf{7.2}
			 & \textbf{9.2}
			 & \textbf{9.3}
			 & 9.2
 \\

\bottomrule

\end{tabular}

\caption{\label{table:augmentation_table} AUC gains $(\Delta \uparrow)$ achieved by a ResNet18 on the corrupted test sets, employing various augmentation strategies. $k$-F refers to the $k$-fold evaluation setting, and the top two strategies are \textbf{bolded}. \textit{Ours} is overall the best performing method ($p<0.05$, Wilcoxon signed-rank test). Note that CutMix and MixUp involve batch-wise transformations and do not natively support multi-label problems. Thus, they do not contribute to the overall mean.}
\end{table}


\section{Conclusions}

This study introduces MedMNIST-C, a comprehensive benchmark specifically designed for assessing algorithm robustness in the context of image corruptions across a wide range of the medical imaging spectrum. We demonstrate the significant advantage of embedding domain knowledge into data augmentation strategies. Our approach, straightforward yet impactful, consistently enhances robustness and outperforms standard methods, even under $k$-fold validation, ensuring an unbiased evaluation against our test corruptions. There is potential to extend our method to simultaneously manage multiple corruptions, mirroring RandAugment, or to implement them across multiple chains as in AugMix. By providing reproducible corrupted test sets and APIs, we aim to encourage the community to both \textit{assess} and \textit{enhance} the robustness of medical image analysis models.

\newpage

\begin{credits}
\subsubsection{\ackname} This study was funded through the Hightech Agenda Bayern (HTA) of the Free State of Bavaria, Germany.
\end{credits}

%
%

%

\bibliographystyle{splncs04}
\bibliography{bibliography.bib}

\end{document}